\documentclass[twoside,11pt]{article}

\usepackage{jmlr2e}

\firstpageno{1}

\usepackage{multirow}

\usepackage{caption}
\usepackage{subcaption}

\newcommand{\E}{\mathbb{E}}
\renewcommand{\Pr}{\mathbb{P}}
\newcommand{\cov}{\mathrm{cov}}
\newcommand{\var}{\mathrm{var}}

\usepackage{datatool}

\begin{document}
\includegraphics[width = .37\linewidth]{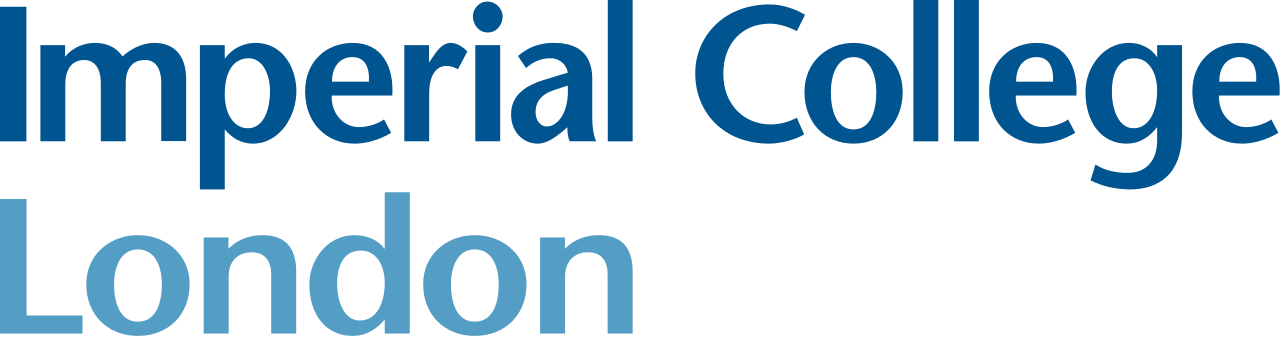}
\vspace*{.2in}

\title{A Bernoulli Mixture Model to Understand and Predict Children's Longitudinal Wheezing Patterns}

\author{\name Pierre G. B. Moutounet-Cartan \email pierre.moutounet-cartan17@imperial.ac.uk \\
       \addr Department of Mathematics\\
       Imperial College London\\
       London, SW7 2AZ, United Kingdom}

\maketitle

\begin{abstract}
\small
In this research, we perform discrete unsupervised machine learning through a Bernoulli Mixture Model on data representing the expression of the wheeze phenotype of patients at different stages of their childhood up to age 16. Wheeze is a distinct noise produced while breathing due to narrowed airways, such as asthma or viral chest infections. Due to a study from \citet{Henderson}, it has been estimated that around 23.5\% of U.K. children had at least wheezed once by six years of age, and 6.9\% had persistent wheezing problems. The usage of a Bernoulli Mixture Model is new in the field, where previous classification methods used classical unsupervised learning such as $K$-means, $K$-medoids, or Latent Class Analyses \citep{Loza, Kurukulaaratchy, Deliu, Brew}. In particular, \citet{Oksel} found that the Latent Class Analysis has resulted majorly dependent to the sample size, and $K$-means is largely dependent on the distance measure and so the data-set.\\

In this research, we estimate that around $27.99(\pm2.15)\%$\footnote{The values in brackets are the differences between the estimated values for the measures listed and the left or right boundary of the 95\% confidence interval. The inverse
cumulative distribution function value is taken from a $t$-distribution with $n-1$ degrees of freedom.} of the population has experienced wheezing before turning 1 in the United Kingdom. Furthermore, the Bernoulli Mixture Model classification is found to work best with $K=4$ clusters in order to better balance the separability of the clusters with their explanatory nature, based on a cohort of $N=1184$. The probability of the group of parents in the $j$th cluster to say that their children have wheezed during the $i$th age is assumed $P_{ij} \sim \text{Beta}(1/2, 1/2)$, the probabilities of assignment to each cluster is $R \sim \text{Dirichlet}_K(\alpha)$, the assignment of the $n$th patient to each cluster is $Z_n\ |\ R \sim \text{Categorical}(R)$, and the $n$th patient wheezed during the $i$th age is $X_{in}\ |\ P_{ij}, Z_n \sim \text{Bernoulli}(P_{i,Z_n})$; where $i\in\{1,\dots,6\}$, $j\in\{1,\dots,K\}$, and $n\in\{1,\dots, N\}$. The classification is then performed through the E-M optimization algorithm \citep{Bishop, Saeed}. We found that this clustering method groups efficiently the patients with late-childhood wheezing, persistent wheezing, early-childhood wheezing, and none or sporadic wheezing. Furthermore, we found that this method is not dependent on the data-set, and can include data-sets with missing entries.\\

It is hoped this study will give medical staff an understanding of the wheezing patterns in children up to age 16, and so provide an effective treatment.
\normalsize
\end{abstract}
\vspace*{.1in}


\tableofcontents

\newpage

\section{The Cohort \& Introduction}

Parents of 1184 children residing in the United Kingdom were asked to tell if they had wheeze in the previous year at ages 1, 3, 5, 8, 11, and 16. A positive answer was recorded as a 1 in the registry, while a negative answer was recorded as a 0. Each patient is given a fixed identification number going from 0 to 1183. The data of 537 children was incomplete (45.4\% of the data), i.e., the parent did not answer to the question at least once. There were 647 full entries, i.e., parents of 647 children provided the information at every stage described above. The information was later put into a spreadsheet, where each row represented the data collected for each child, and each column the answer for each of the periods as explained above, under "Age 1," "Age 3," "Age 5," "Age 8," "Age 11," and "Age 16." Whenever no information was provided, NaN was recorded.\\

From this data, we can therefore estimate the number of children within the U.K. population who had wheeze at most a year before turning 1, 3, 5, 8, 11, and 16. We estimate the means by the sample means, and produce a 95\% confidence interval with the sample mean and the corresponding critical value from a $t$-distribution with 647 degrees of freedom. This algorithm can be found in Appendix.\\

\begin{table}[h!]
    \centering
    \begin{tabular}{|c||c|cc|}
       \hline Age period  & Est. mean & LHS CI at 95\% & RHS CI at 95\% \\\hline\hline
Age 1   &      27.99\%     &       25.84\%     &       30.14\% \\
Age 3    &     23.74\%     &       21.70\%     &       25.78\% \\
Age 5   &      23.04\%     &       21.02\%     &       25.06\% \\
Age 8   &      18.05\%     &       16.20\%     &       19.89\% \\
Age 11  &      18.89\%     &       17.01\%     &       20.76\% \\
Age 16  &      17.02\%     &       15.22\%     &       18.83\%  \\\hline
    \end{tabular}
    \caption{Estimate of the U.K. population having wheezed at different ages with the LHS and RHS of the 95\% confidence interval using a $t$-distribution with 647 degrees of freedom.}
    \label{tabmeans}
\end{table}

From Table \ref{tabmeans}, we can deduce that approximately $27.99(\pm2.15)\%$ (with a 95\% confidence) of the population has experienced wheezing before turning 1. This seems to corroborate \citet{Henderson} findings, where they estimated that 23.5\% of U.K. children had at least wheezed once by 6.\\

\begin{figure}[h!]
    \centering
    \includegraphics[width=0.8\linewidth]{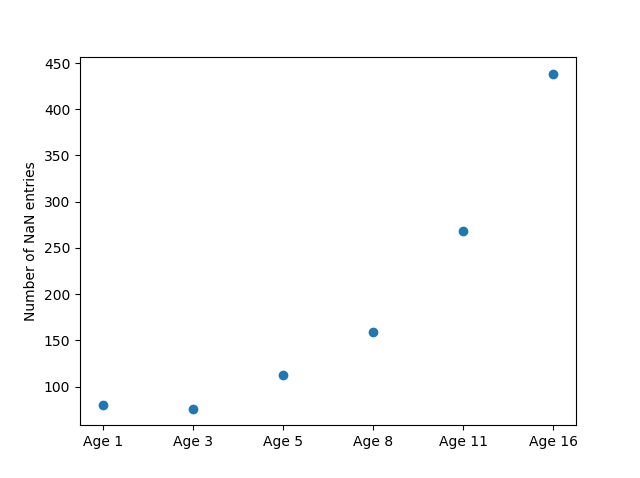}
    \caption{Number of NaN entries per age tranche within the cohort data.}
    \label{nanvalue}
\end{figure}

The challenge with this data is that 45.5\% of it is incomplete. From Fig. \ref{nanvalue}, the number of parents stopping to provide a status on their child wheeziness increases whenever the latter gets older.

In particular, other papers such as \citet{Loza, Kurukulaaratchy, Deliu, Brew, Oksel} have used well known clustering methods such as K-means, K-medoids, as well as Latent Class Analysis. These methods require that no entry be missing or classified as NaN in the data-set, which we have for a large proportion in our case. Furthermore, \citet{Oksel} found that the Latent Class Analysis has resulted majorly dependent to the sample size, the frequency, and the timing of data collection.\\

Therefore, in this paper, we suggest a novel method for classifying the longitudinal wheeze phenotype diagnosed in children aged 1 to 16, using a Bernoulli Mixture Model based on the E-M optimization algorithm. This method can allow missing entries in the data-set, and is generally less dependent on the sample size than other methods.

\section{Theory of Bernoulli Mixture Model}

In this subsection, we consider the general case of the Bernoulli Mixture Model so that we can apply it in our case later on.\\

Let $X_1,\dots,X_M$ be a set of $M$ binary variables where $X_i\sim\mathrm{Bernoulli}(\lambda_i)$ for $\lambda_i\in(0,1)$, $\forall i\in\{1,2,\dots,M\}$. Note that in this case, the marginals of $X_i$ are given by $f_{X_i}(x_i)=\lambda_i^{x_i}(1-\lambda_i)^{1-x_i}$ where $x_i\in\{0,1\}$, $\forall i\in\{1,2,\dots,M\}$.\\ 

Hence, we have that
\[\Pr(\mathbf{X}|\Lambda) = \prod_{i=1}^M f_{X_i}(x_i) = \prod_{i=1}^M \lambda_i^{x_i}(1-\lambda_i)^{1-x_i}\]
where $\mathbf{X}=(X_1,\dots,X_M)^t$ and $\Lambda=(\lambda_1,\dots,\lambda_M)^t$.\\

As the above expression are written as a separate product of the different probability mass functions of $X_1,\dots,X_M$, we have that $\mathbf{X}$ is independent of $\Lambda$. We have $\E(X_i|\lambda_i)=\lambda_i, \forall i\in\{1,\dots,M\}$ by characteristic of the Bernoulli distribution, so that
\[\E(\mathbf{X}|\Lambda) = \Lambda\in(0,1)^M\ldotp\ (\dagger)\]
Similarly, by independence we have $\cov(X_i,X_j|\lambda_i,\lambda_j)=0$, $\forall i,j\in\{1,\dots,M\}$ such that $i\neq j$, and by property of the Bernoulli distribution, $\var(X_i|\lambda_i)=\lambda_i(1-\lambda_i)$, $\forall i\in\{1,\dots,M\}$. Therefore,
\[\mathbf{\Sigma}_{ik}:=\cov(\mathbf{X}|\Lambda)_{ik} = \delta_{ik} \lambda_i(1-\lambda_i)\in(0,1)^{M\times M}\ (\ddagger)\]
where $\delta_{ik}$ is the Kronecker delta, i.e.,
\[\delta_{ik}=\begin{cases}1\text{ if }i=k,\\ 0\text{ if }i\neq k\ldotp\end{cases}\]

We now build a finite mixture, with new parameters $\Pi = (\pi_1,\dots,\pi_K)^t$, where $K\leqslant M$ will be the number of clusters, by
\[\Pr(\mathbf{X}|\tilde{\Lambda},\Pi) := \sum_{j=1}^K \pi_j\Pr(\mathbf{X}|\Lambda_j)\ (\star)\]
where $\tilde{\Lambda}=(\Lambda_1,\dots,\Lambda_K)^t$, the set of parameters of each component \citep{Saeed}, where we have
\[\Pr(\mathbf{X}|\Lambda_j) = \prod_{i=1}^M \lambda_{ji}^{x_i}(1-\lambda_{ji})^{1-x_i}\ldotp\]

For this mixture model, we have
\[\E(\mathbf{X}|\tilde{\Lambda},\Pi) = \sum_{j=1}^K \pi_i\E(\mathbf{X}|\Lambda_j) = \sum_{j=1}^K \pi_j\Lambda_j\in(0,1)^M\text{ by }(\dagger)\ldotp\]

Similarly,
\[\cov(\mathbf{X}|\tilde{\Lambda},\Pi) = -\E(\mathbf{X}|\tilde{\Lambda},\Pi)\E(\mathbf{X}|\tilde{\Lambda},\Pi)^t+\sum_{j=1}^K\pi_k(\mathbf{\Sigma}_j+\Lambda_j\Lambda_j^t)\]
where $\mathbf{\Sigma}_j$ is as defined in $(\ddagger)$, i.e., 
\[(\mathbf{\Sigma}_j)_{ik} = \delta_{ik}\lambda_{ji}(1-\lambda_{ji})\ldotp\]

For a random sample $\mathbf{X}_1,\dots,\mathbf{X}_N$ which is distributed as in $(\star)$ with respect to $\tilde{\Lambda}$ and $\Pi$, the likelihood function for $\mathbf{X}_1,\dots,\mathbf{X}_N|\tilde{\Lambda},\Pi$ is given by
\[\mathcal{L}(\mathbf{X}_1,\dots,\mathbf{X}_N,\tilde{\Lambda},\Pi) = \prod_{n=1}^N\sum_{j=1}^K \pi_j\Pr(\mathbf{X}_n|\Lambda_j)\]
so that the log-likelihood $\ell$ is
\[\ell(\mathbf{X}_1,\dots,\mathbf{X}_N,\tilde{\Lambda},\Pi) =\sum_{n=1}^N\log\left(\sum_{j=1}^K\pi_j\Pr(\mathbf{X}_n|\Lambda_j)\right)= \sum_{n=1}^N\log\left(\sum_{j=1}^K\left\{\pi_j\prod_{i=1}^M \lambda_{ji}^{x_{ni}}(1-\lambda_{ji})^{1-x_{ni}}\right\}\right)\ldotp\]

The aim is to generate an algorithm that finds $\Pi$, $\tilde{\Lambda}$ which maximize $\ell$, which we will call $\Pi_{\text{max}}$ and $\tilde{\Lambda}_{\text{max}}$ respectively. Unfortunately, due to the shape of $\ell$, $\Pi_{\text{max}}$ and $\tilde{\Lambda}_{\text{max}}$ cannot be found in closed form in that case. \\

Hence, we introduce a latent binary variable $\mathbf{Z}=(z_1,\dots,z_K)^t$ associated with each instance of $\mathbf{X}$ \citep{Bishop}. Therefore, the conditional distribution of $\mathbf{X}$, given the latent variable $\mathbf{Z}$, is given by
\[\Pr(\mathbf{X}|\mathbf{Z}, \tilde{\Lambda}) = \prod_{j=1}^K \Pr(\mathbf{X}|\Lambda_j)^{z_j},\]
where the prior for the latent variable is 
\[\Pr(\mathbf{Z}|\Pi) = \prod_{j=1}^K \pi_j^{z_j}\ldotp\]

By considering the random sample $\mathbf{X}_1,\dots,\mathbf{X}_N$, we have the likelihood function $\mathcal{G}$ for $\mathbf{X}_1,\dots,\mathbf{X}_N|\mathbf{Z}$ given by
\[\mathcal{G}(\mathbf{X}_1,\dots,\mathbf{X}_N,\mathbf{Z},\tilde{\Lambda},\Pi) = \prod_{n=1}^N \Pr(\mathbf{X}_n,\mathbf{Z}|\tilde{\Lambda},\Pi) = \prod_{n=1}^N\prod_{j=1}^K (\pi_j\Pr(\mathbf{X}_n|\Lambda_j))^{z_{nj}},\]
so that, once expanded, we have
\[\mathcal{G}(\mathbf{X}_1,\dots,\mathbf{X}_N,\mathbf{Z},\tilde{\Lambda},\Pi) = \prod_{n=1}^N\prod_{j=1}^K\left(\pi_j\prod_{i=1}^M\lambda_{ji}^{x_{ni}}(1-\lambda_{ji})^{1-x_{ni}}\right)^{z_{nj}}\ldotp\]

Therefore, the log-likelihood $g$ is given by
\[g(\mathbf{X}_1,\dots,\mathbf{X}_N,\mathbf{Z},\tilde{\Lambda},\Pi) = \sum_{n=1}^N\sum_{j=1}^Kz_{nj}\left(\log(\pi_j)+\sum_{i=1}^M\left\{ x_{ni}\log(\lambda_{ji})+(1-x_{ni})\log(\lambda_{ji})\right\}\right)\ldotp\]

Hence, the expectation of the log-likelihood with respect to the marginal distribution of $\mathbf{Z}$ is given by
\[\E_{\Pr(\mathbf{Z}|\Pi)}(g(\mathbf{X}_1,\dots,\mathbf{X}_N,\mathbf{Z},\tilde{\Lambda},\Pi)) = \sum_{n=1}^N\sum_{j=1}^K\E(z_{nj})\left(\log(\pi_j)+\sum_{i=1}^M\left\{ x_{ni}\log(\lambda_{ji})+(1-x_{ni})\log(\lambda_{ji})\right\}\right)\ldotp (\star\star)\]

By Bayes' Theorem and \citet{Saeed, Bishop}, we have
\[\E(z_{nj}) = \dfrac{\pi_j \Pr(\mathbf{X}_n|\Lambda_j)}{\sum_{k=1}^K \pi_k\Pr(\mathbf{X}_n|\Lambda_k)}\ldotp\ (\diamond)\]

Let 
\[N_j = \sum_{n=1}^N \E(z_{nj})\text{ and }\mathbf{\overline{X}}_j = \dfrac{1}{N_j}\sum_{n=1}^N\E(z_{nj})\mathbf{X}_n\ldotp\]

Then one can show that
\[\tilde{\Lambda}_{\text{max}} = (\hat{\Lambda}_1,\dots,\hat{\Lambda}_K)^t \text{ where }\hat{\Lambda}_j = \mathbf{\overline{X}}_j\]
\[\text{ and }\hat{\Pi}_{\text{max}} = (\hat{\pi}_1,\dots,\hat{\pi}_K)^t\text{ where }\hat{\pi}_j = N_j/N\ldotp\]

Indeed, such a $\tilde{\Lambda}_{\text{max}}$ makes the derivative with respect to the $\lambda_j$s vanish, and such a $\hat{\Pi}_{\text{max}}$ maximizes $(\star\star)$ through a Lagrange multiplier as seen in \citet{Saeed, Bishop}.\\

Therefore, the expectation-maximization algorithm for a Bernoulli mixture model first gives initialization values to $\tilde{\Lambda}$ and $\Pi$. The algorithm then computes the value of the log-likelihood at the initial values $\tilde{\Lambda}_0$, $\Pi_0$.\\

On the next step, the algorithm does a loop by evaluating $\E(z_{nj})$ as in $(\diamond)$, and reevaluates $\hat{\Lambda}_j, \hat{\pi}_j$ as found above. Then, we evaluate the log-likelihood at these values. We stop the loop whenever the log-likelihood meets a convergence criterion.

\section{Application to the data-set}

Each cluster has a certain probability of a yes answer at the different ages of the children. We assume that they follow a Beta distribution. Hence, let
\[P_{ij} \sim \mathrm{Beta}\left(\dfrac{1}{2},\dfrac{1}{2}\right)\text{ for }i\in\{1,\dots,6\}\text{ and} j\in\{1,\dots,K\}\]
where $K$ is the number of clusters, as the Beta distribution is well known to represent a distribution of probabilities, and $\mathrm{Beta}(1/2,1/2)$ is more dense around its support boundaries $0$ and $1$. Hence, this is our prior. That is, given a cluster $j\in\{1,\dots,K\}$ and the age tranche of the children $i\in\{1,\dots,6\}$, the group of parents in the $j$th cluster has a probability of $P_{ij}$ to say that their children have wheezed at (or within a year of) the $i$th age.\\

Now, to provide the cluster assignments, we use the Categorical distribution, which is a generalized Bernoulli distribution (in higher dimensions), as each vector in the data either belongs to one of the $K$ clusters or does not. Hence, let
\[Z_n\ |\ R \sim \mathrm{Categorical}_K(R)\text{ for } n\in\{1,\dots,N\},\]
where $R$ represents the probabilities of the cluster assignments, which we assume follows a Dirichlet distribution, i.e., a generalized Beta distribution (in higher dimensions), so that
\[R\sim \mathrm{Dirichlet}_K(\alpha)\]
for some constant and positive vector $\alpha$, which we are free to choose as it is an uninformative distribution.\\

Therefore, for $X_{in}$ the random variable representing the fact that the $n$th patient wheezed at the $i$th age tranche, we have
\[X_{in}\ |\ P_{ij}, Z_n \sim \mathrm{Bernoulli}(P_{i,Z_n})\text{ for }i\in\{1,\dots,6\}\text{ and }n\in\{1,\dots,N\}\ldotp\]

To sum up, the model is given by
    \begin{align*}
        P_{ij} &\sim \text{Beta}(1/2,1/2),\\
        R &\sim \text{Dirichlet}_K(\alpha),\\
        Z_n\ |\ R &\sim \text{Categorical}_K(R),\\
        X_{in}\ |\ P_{ij}, Z_n &\sim \text{Bernoulli}(P_{i,Z_n}),
    \end{align*}
    for $i\in\{1,\dots,6\}$, $j\in\{1,\dots,K\}$, and $n\in\{1,\dots, N\}$, where each variable describes the following: \begin{align*}
        P_{ij}&\text{: probability of the group of parents in the }j\text{th cluster to say that their children have}\\
        &\hspace{.2in} \text{wheezed during the }i\text{th age,}\\
        R&\text{: probabilities of assignment to each cluster }1,...,K,\\
        Z_n&\text{: assignment of the }n\text{th patient to each cluster } 1,...,K, \\
        X_{in}&\text{: the }n\text{th patient wheezed during the }i\text{th age.}
    \end{align*}
    
\begin{figure}[h!]
    \centering
    \includegraphics[width = .75\linewidth]{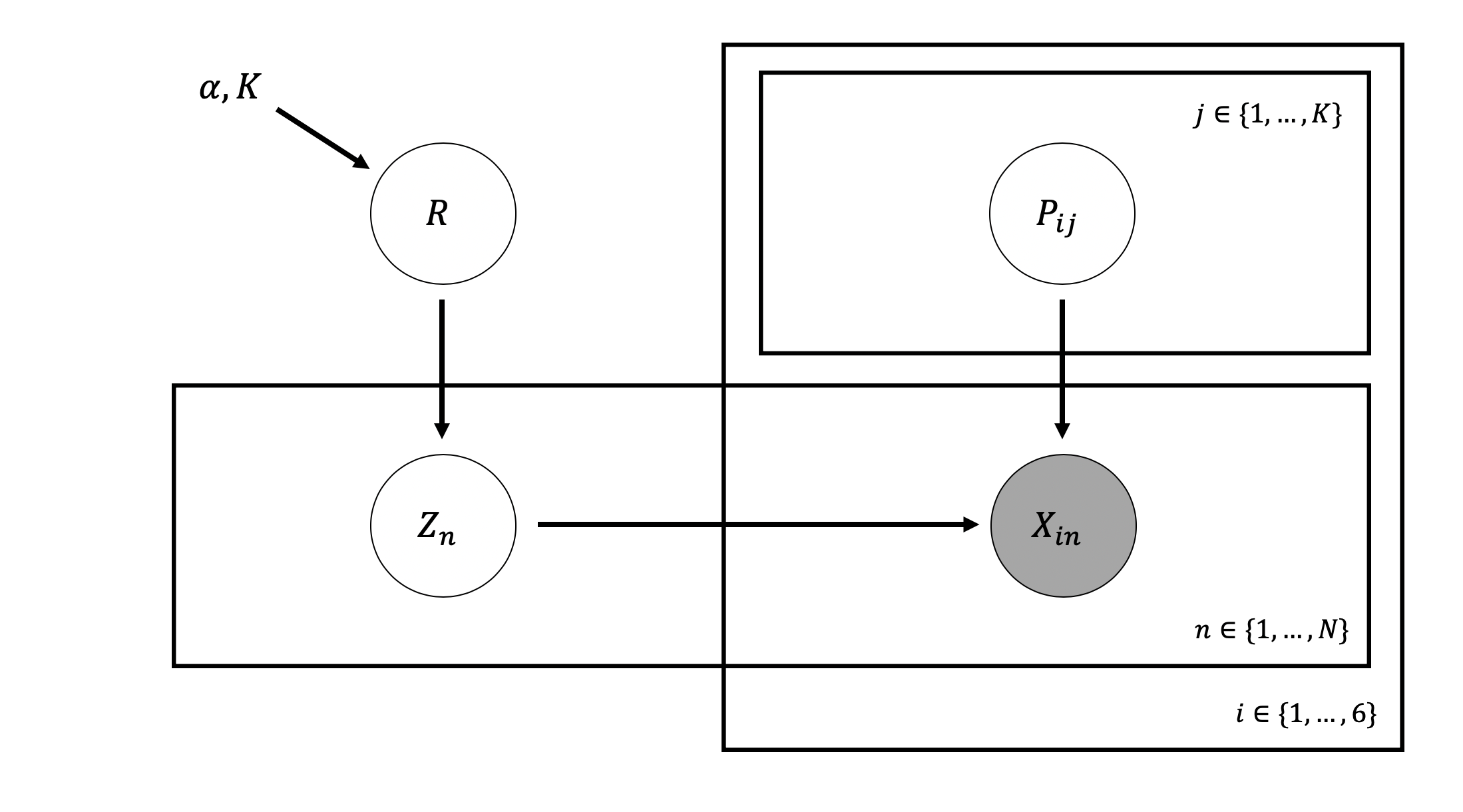}
    \caption{Plate diagram of the considered Bernoulli Mixture Model for the considered cohort.}
    \label{fig:platediagram}
\end{figure}

The plate diagram for such a configuration can be found in Fig. \ref{fig:platediagram}. Note again that $\alpha$, a $K$-dimensional vector with $\alpha_j>0$ for all $j\in\{1,\dots,K\}$, can be chosen arbitrarily since the distribution of $R$ is uninformative ($Z_n$ is latent).\\
    
\section{Results}

\begin{figure*}[h!]
    \centering
    \begin{subfigure}[c]{0.5\textwidth}
        \centering
        \includegraphics[width=0.9\linewidth]{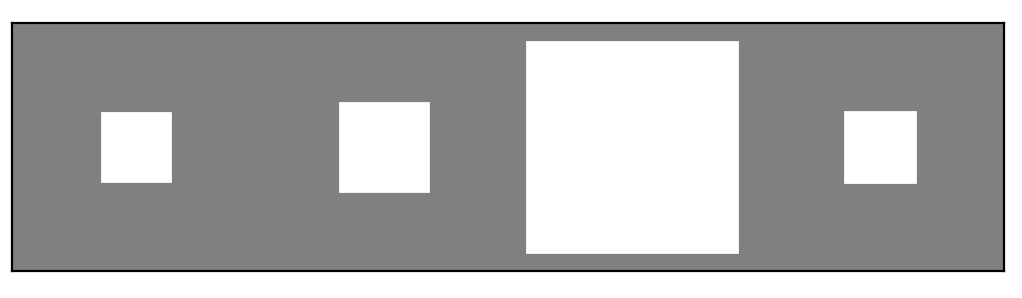}
        \caption{The Hinton plot for $4$ clusters of $R$.}
    \end{subfigure}%
    ~ 
    \begin{subfigure}[c]{0.5\textwidth}
        \centering
        \includegraphics[width=0.9\linewidth]{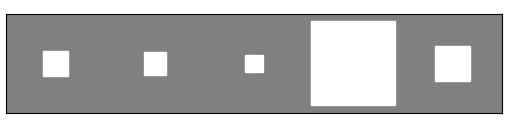}
        \caption{The Hinton plot for $5$ clusters of $R$.}
    \end{subfigure}
    
    \begin{subfigure}[c]{0.5\textwidth}
        \centering
        \includegraphics[width=0.9\linewidth]{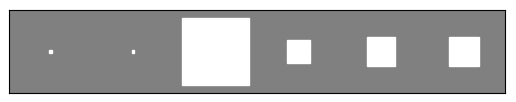}
        \caption{The Hinton plot for $6$ clusters of $R$.}
    \end{subfigure}%
    ~ 
    \begin{subfigure}[c]{0.5\textwidth}
        \centering
        \includegraphics[width=0.9\linewidth]{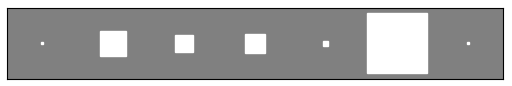}
        \caption{The Hinton plot for $7$ clusters of $R$.}
    \end{subfigure}
    \caption{The Hinton plots of $R$ for different number of clusters. Each graph shows the probabilities of being assigned to each cluster. Here, the data used excludes the rows with NaN entries.}
    \label{HintonR}
\end{figure*}

We then perform the expectation-maximization algorithm on the log-likelihood as explained in the previous section. We can then plot the Hinton plots for $R$, the $P_{ij}$s, and the $Z_n$s. These are visualized in Fig. \ref{HintonR}.\\

\begin{figure}[h!]
    \centering
    \includegraphics[width=1.\linewidth]{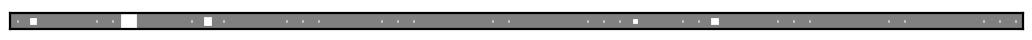}
    \caption{The Hinton diagram by setting $64$ clusters of $R$.}
    \label{Hintonmax}
\end{figure}

Hinton diagrams for $R$ show the number of elements per cluster. The bigger the white square is, the bigger the cluster is.  From Fig. \ref{HintonR}, we can see that for either setting $4,5,6$ or $7$ groups, there are $4$ dominant clusters. Now, if we set the number of clusters to $64=2^6$, which is the number of all possibilities to arrange 0s and 1s in a 6-dimensional vector, i.e., all ways parents can answer to the questionnaire, then we get the Hinton diagram for $R$ as shown in Fig. \ref{Hintonmax}.\\

\begin{figure}[h!]
    \centering
    \includegraphics[width=0.45\linewidth, angle = -90]{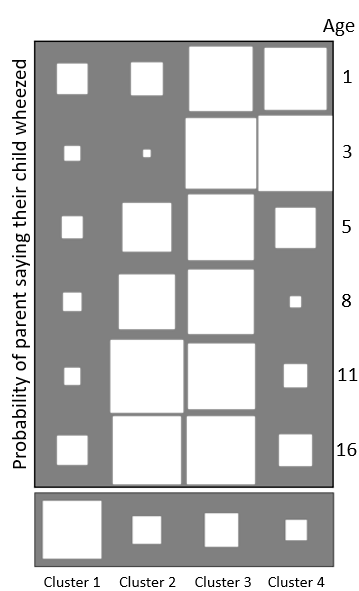}
    \caption{Hinton diagram of the $P_{ij}$s for $i\in\{1,\dots,6\}$ and $j\in\{1,\dots,4\}$, i.e., the probabilities of the parents saying that their child wheezed at each age tranche separated by cluster. The squares in the lower rectangle show the "sizes" of each cluster (more formally, the probability of belonging to each cluster).}
   \label{HintonProbs}
\end{figure}

\begin{figure}[h!]
     \centering
     \begin{subfigure}[b]{.49\textwidth}
         \centering
         \includegraphics[width = .99\linewidth]{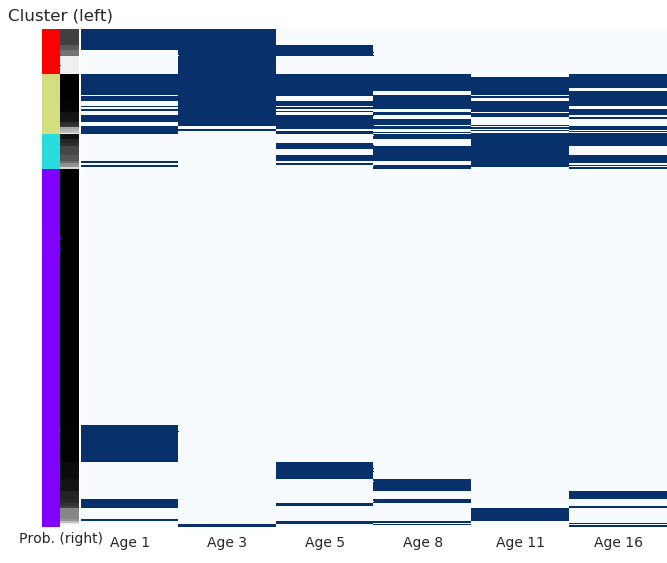}
         \caption{Rows with missing entries disregarded.}
         \label{BernoulliHeat}
     \end{subfigure}
     \hfill
     \begin{subfigure}[b]{.49\textwidth}
         \centering
         \includegraphics[width = .99\textwidth]{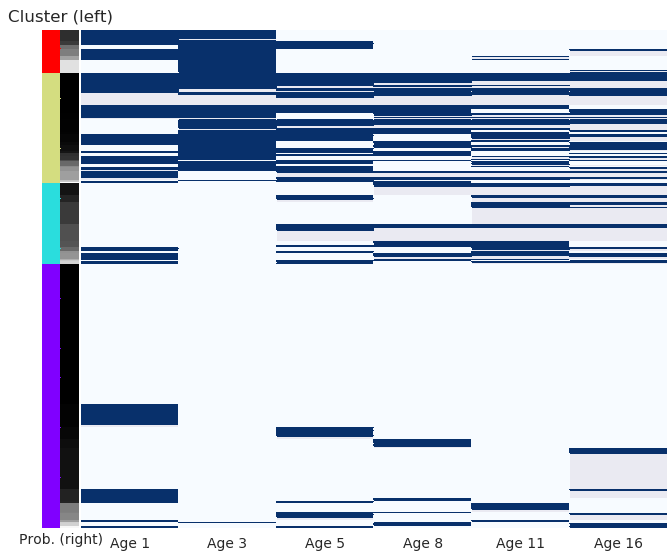}
         \caption{Rows with missing entries kept.}
         \label{BernoulliHeatall}
     \end{subfigure}
    \caption{Heat map of the answers of the parents and their respective cluster (the cluster they have been assigned with highest probability) where (\ref{BernoulliHeat}) the rows with missing entries were disregarded, and where (\ref{BernoulliHeatall}) the rows with missing entries were all kept. Each dark blue entry represents a positive answer by the parents that their child had wheezed, while a light blue entry is a negative answer. From Fig. \ref{HintonProbs}, Cluster 1 is here the purple group, Cluster 2 is the blue group, Cluster 3 is the green group and Cluster 4 is the red group. The gray-scaled bar represents the probability of assignment to this cluster, going from $0.4$ (white) to $1$ (black).  Each missing answer for Fig. \ref{BernoulliHeatall} from the parents is shown in gray.}
   \label{BernoulliHeat_both}
\end{figure}

We can see from Fig. \ref{Hintonmax} that there are $4$ dominant clusters. Hence, later on for the Bernoulli mixture model we will take 4 different clusters. Note that under the implemented algorithm, although the clusters -- of course -- do not change by launching again the code, the order of these clusters do change. A choice of 4 clusters give the following Hinton diagram for the $P_{ij}$s, as seen in Fig. \ref{HintonProbs}, superimposed with the Hinton diagram for $R$. According to Fig. \ref{HintonProbs}, Cluster 2 is made of the children with high probabilities of late wheezing, Cluster 4 of the children with high probabilities of early wheezing, Cluster 3 of the children with high probabilities of persisting wheezing, and Cluster 1 of the children with probabilities of sporadic and benign wheezing. To give ourselves an idea of the assignments of each cluster for each patient, we can plot the cluster heat map as shown in Fig. \ref{BernoulliHeat}. As seen in Fig. \ref{BernoulliHeat}, this clustering method seems to be grouping quite efficiently the patients with late-childhood wheezing (blue group), persistent wheezing (green group), early-childhood wheezing (red group), and none or sporadic wheezing (purple group). \\

The answers in each group that look more like outliers do have a lower probability to be assigned this cluster, as seen classified in Fig. \ref{BernoulliHeat}. For example, the set of answers (No, No, Yes, No, No, Yes) belongs to the purple cluster with smallest probability of assignment in this cluster, at approximately 46\% (but still higher than the probabilities of assignment in other clusters). The highest in the purple cluster being 97\% for the set of answers (No, No, No, No, No, No). For the blue cluster, the smallest assignment probability is approximately 51\% with set of answers (Yes, No, Yes, No, Yes, Yes) while the highest is 97\% with answers (No, No, No, Yes, Yes, Yes) -- which makes sense as it is the cluster of late-wheezing children. For the green cluster, the smallest is 51\% with answers (Yes, No, Yes, Yes, No, No), the highest being at 99.9\% with answers (Yes, Yes, Yes, Yes, Yes, Yes) since it is the cluster of persistent wheeze issues children. Finally, for the red cluster, the smallest probability is 61\% for the answers (No, Yes, No, No, No, Yes), the highest being 98\% for the answers (Yes, Yes, No, No, No, No) as it is the cluster of early-wheezing children.\\

\begin{figure}[h!]
    \centering
    \includegraphics[width=.7\linewidth]{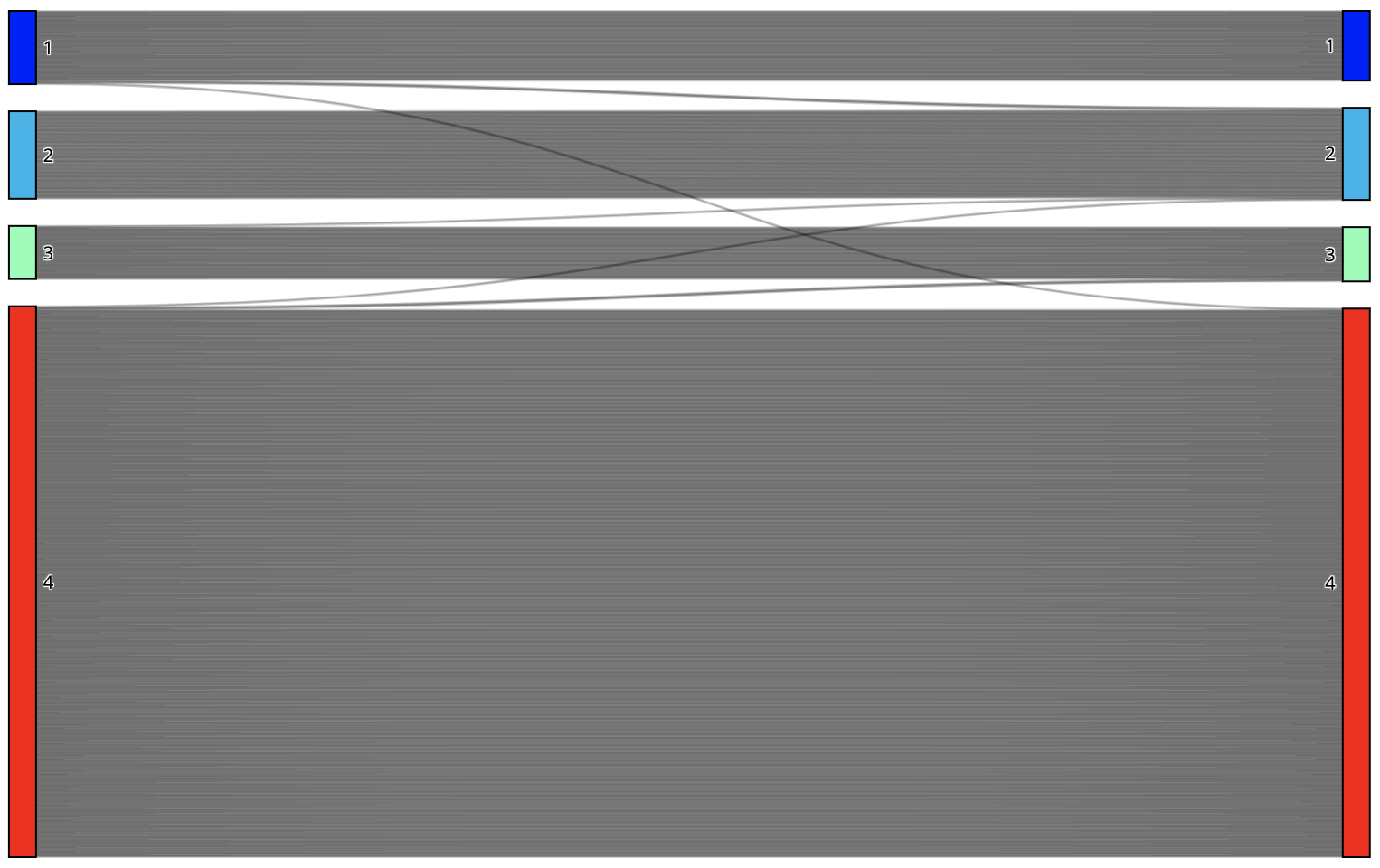}
    \caption{Sankey diagram showing the patients flow between Bernoulli mixture models with the complete data only (left) and all the data including rows with missing entries (right). Note that the clusters here may not have the same labels and color as before since the algorithm was called another time here.}
    \label{SankeyBig}
\end{figure}

We can also perform the Bernoulli process on the full data, i.e., also on the parents that did not say if their child wheezed at all age tranches, since $P_{ij}$ is a probability assigned to each answer individually. The aim is to compare whether the change of grouping of the data with answers for each age tranche changes whether or not we add up this new incomplete data. Ideally, it does not so that the prediction is accurate. The heat map of the clustering via Bernoulli mixture of all the data, including the missing one, is shown in Fig.~\ref{BernoulliHeatall}.\\

Comparing the sizes of the clusters in Fig. \ref{BernoulliHeat} and Fig. \ref{BernoulliHeatall}, adding the data with missing entries has increased the sizes of the green and blue clusters. We can here plot a Sankey diagram, as seen in Fig. \ref{SankeyBig} showing to which clusters the entries with complete data are mapped from a Bernoulli mixture with only complete data to a Bernoulli mixture with all the data available. From Fig. \ref{SankeyBig}, we can see that the model for full entries remains stable with only seven swing patients, of which one of type $\alpha$ (Yes, No, Yes, No, Yes, Yes), two of type $\beta$ (Yes, Yes, No, No, Yes, No), one of type $\gamma$ (Yes, No, Yes, No, No, Yes), two of type $\xi$ (No, No, No, Yes, No, Yes), and one of type $\lambda$ (No, Yes, No, No, No, Yes). \\

Those swing patients are shown in Table \ref{tableSwings}. The "biggest" swing is made by the two patients of type $\beta$ (see Table \ref{tableSwings}), where they have above 60\% of probability to be assigned different clusters when the missing entries from the full set of data is added up. This is due to the fact that the expression of the wheeze phenotype of these patients is very unpredictable and do not show any pattern -- the parents answered (Yes, Yes, No, No, Yes, No). 

\begin{table}[h!]
    \centering
    \begin{tabular}{|c|cc|cc|}
    \hline
    \textsc{Patient type} & \textsc{Cl L} & \textsc{Pr L} & \textsc{Cl R} & \textsc{Pr R}\\\hline\hline
    $\alpha$  & 3 & 51\% & 2 & 55\% \\
    $\beta$ & 1 & 67\% & 2 & 65\% \\
    $\gamma$ & 4 & 64\% & 2 & 40\% \\
    $\xi$ & 4 & 56\% & 3 & 57\% \\
    $\lambda$ & 1 & 41\% & 4 & 54\% \\
    \hline
    \end{tabular}
    \caption{Allocations of the swing patients from a Bernoulli mixture with only the clean data (assigned to the cluster \textsc{Cl L} with probability \textsc{Pr L}) to all the data (assigned to the cluster \textsc{Cl R} with probability \textsc{Pr R}), where each cluster number is as shown in Fig. \ref{SankeyBig}.}
    \label{tableSwings}
\end{table}

The Bernoulli Mixture Model is advantageous in practice because it allows practitioners to add up data throughout the childhood of a patient. E.g., if the parents have a child aged 3 and gave answers to the questionnaire (whether or not the child wheezed before 1, and around 3), the data can still be clustered through the Bernoulli Mixture Model, so that the practitioner can look at complete past entries in this group ("neighboring data") and predict future wheezing patterns of this child. As seen in Fig. \ref{BernoulliHeatall}, the Bernoulli Mixture Model generally assumes higher probabilities of wheeze if no entry is given, unless if it is sporadic. This method might be more accurate if the parents are asked if their children wheezed during smaller age intervals, such as yearly. \\

\newpage

\section{Conflicts of interest}

The author(s) note no known conflict of interest by undertaking this research.

\newpage

\vskip .2in
\bibliography{sample}

\end{document}